\documentclass[twocolumn,floats,floatfix,aps,pra]{revtex4}
\usepackage{eurosym}
\usepackage{amsfonts}
\usepackage{amssymb}
\usepackage{amsmath}
\usepackage{graphicx}
\usepackage{bm}
\usepackage{color}
\usepackage{epsfig}
\usepackage{ifthen}
\usepackage{subfigure}
\usepackage{physics}
\usepackage[hidelinks, colorlinks=true,linkcolor=blue,citecolor=blue,filecolor=blue,urlcolor=blue]{hyperref}

\begin{document}

\title{Creation of coherent superpositions of Raman qubits by using dissipation}
\author{Andon A. Rangelov}
\affiliation{Center for Quantum Technologies, Department of Physics, Sofia University, James Bourchier 5 blvd., 1164
Sofia, Bulgaria}
\author{Nikolay V. Vitanov}
\affiliation{Center for Quantum Technologies, Department of Physics, Sofia University, James Bourchier 5 blvd., 1164
Sofia, Bulgaria}

\begin{abstract}
We show how to create coherent superpositions between two ground states of $\Lambda$ quantum system of three states, among which the middle one decays. 
The idea is to deplete the population of the bright state formed by the two ground states via the population loss channel. 
The remaining population is trapped in the dark states, which can be designed to be equal to any desired coherent superposition of the ground states.
The present concept is an alternative to the slow adiabatic creation of coherent superpositions and may therefore be realized over short times, especially in the case where the middle state has a short life span. 
However, the price we pay for the fast evolution is associated with an overall 50\% population losses.
This issue can be removed in an experiment by using post-selection.
\end{abstract}

\author{}
\maketitle

\section{Introduction}

A well-established, effective and robust method for complete population transfer or creation of coherent superpositions between the two ground states of three-state quantum system in $\Lambda$ configuration is stimulated Raman adiabatic passage (STIRAP) \cite{Gaubatz, Vitanov2001A,Vitanov2001B,Vitanov2017,Unanyan1998,Vitanov1999}. 
For the case of complete population transfer two partially overlapping delayed laser pulses in the counterintuitive order transfer adiabatically the population between states $\left\vert 1\right\rangle $ and $\left\vert 3\right\rangle $ via an intermediate state $\left\vert 2\right\rangle$. 
While the intermediate state $\left\vert 2\right\rangle $ can be off-resonance by a certain detuning, the starting state $\left\vert 1\right\rangle $ and the end state $\left\vert 3\right\rangle $ must be on two-photon resonance. 
A dark eigenstate, corresponding to eigenvalue zero of the Hamiltonian, which is a linear superposition of the bare states $\left\vert 1\right\rangle $ and $\left\vert 3\right\rangle $ alone, is used to transfer the population. Since no population ever exists in the state $\left\vert 2\right\rangle $ in the adiabatic limit, its characteristics have no effect on the transfer efficiency. 
In order to create coherent superpositions, the Stokes pulse arrives before the pump pulse but unlike traditional STIRAP, the two pulses terminate simultaneously while maintaining a constant ratio of amplitudes \cite{Vitanov1999}. 
However, for complete population transfer as well as for the creation of coherent superpositions, STIRAP requires very long pulses in order to have a sufficiently large pulse area and thus to fulfill the adiabatic conditions \cite{Vitanov2001A,Vitanov2001B,Vitanov2017}.
Furthermore, beyond the adiabatic limit, i.e., for finite pulse areas, the intermediate state is populated both during and after the excitation and the efficiency of the transfer process drops due to dissipation from the intermediate state \cite{Vitanov1997}.

In this paper, we describe a technique for the creation of coherent superpositions between the two ground states of three-state quantum system in the $\Lambda$ configuration. 
The present approach uses a decaying intermediate state as a key feature to the evolution of the system and the creation of superpositions.
It is shown that the above mechanism is very efficient and robust. 
It works at a short time scale provided that the decay losses from the middle state are large enough. 
Therefore in contrast to STIRAP, here very fast evolution is possible in the case where the middle state is highly lossy.
%\textit{Therefore in contrast to STIRAP, here very fast evolution is possible in the case where the middle state is highly lossy.}
%\textcolor{red}{\textit{NV: Not sure about being faster. Large decay means many decays during process duration, which means large duration.}}
The present approach may be considered as an alternative to STIRAP in situations when the middle state decay is too large: instead of trying to avoid it by making STIRAP more adiabatic, here it is used deliberately, for good.

\section{The three-state system}

\begin{figure}[tb]
\centerline{\includegraphics[width=0.5\columnwidth]{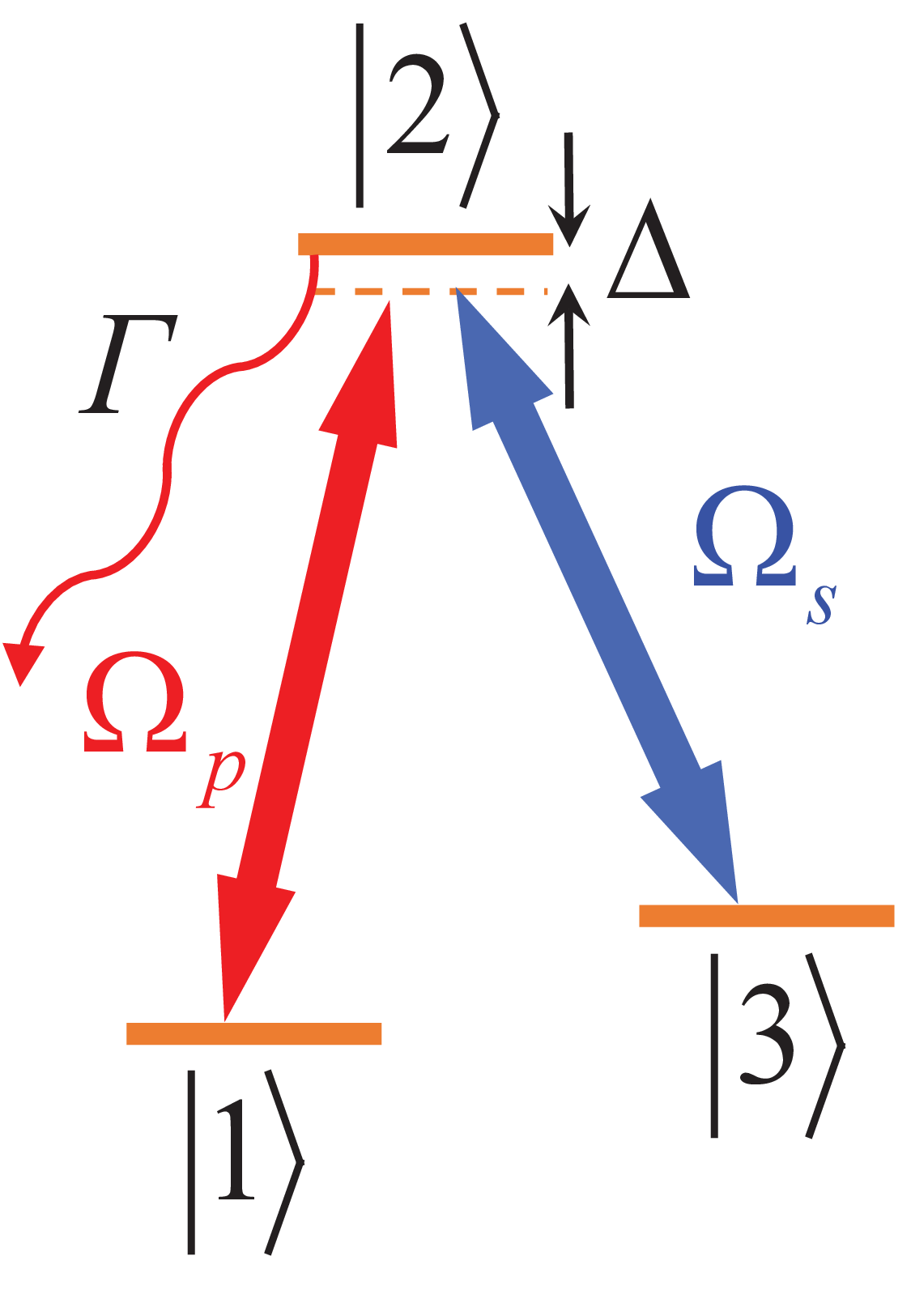}}
\caption{(Color online) The three-state $\Lambda $ system under
consideration. The laser pulse $\Omega _{p}$ couples states $\left\vert
1\right\rangle $ and $\left\vert 2\right\rangle $, whereas the laser pulse $%
\Omega _{s}$ couples states $\left\vert 2\right\rangle $ and $\left\vert
3\right\rangle $. The direct transitions between $\left\vert 1\right\rangle $
and $\left\vert 3\right\rangle $ is prohibited, but the states are on
two-photon resonance. With a total decay rate of $\Gamma $, the intermediate
state $\left\vert 2\right\rangle $ decays out of the system after being off-resonance by a detuning of $\Delta $.}
\label{fig1}
\end{figure}

Figure \ref{fig1} presents a schematic representation of the three-state $\Lambda $ system under study.
The pump-laser pulse $\Omega _{p}(t)$ couples states $\left\vert 1\right\rangle $ and $\left\vert 2\right\rangle $, while the Stokes-laser pulse $\Omega _{s}(t)$ couples states $\left\vert 2\right\rangle $ and $\left\vert 3\right\rangle $. 
It is forbidden by the electric-dipole transition to go directly from state $\left\vert 1\right\rangle $ to state $\left\vert 3\right\rangle $. 
There is maintenance of two-photon resonance between states $\left\vert 1\right\rangle $ and $\left\vert 3\right\rangle $. 
The intermediate state $\left\vert 2\right\rangle $ decays out of the system via a specific process, e.g., spontaneous emission, ionization, (deliberate) coupling to an outside state, etc. with a total decay rate $\Gamma >0$. 
State $\left\vert 2\right\rangle $ may be off-resonance by a detuning $\Delta $.

%***************************************************************
%***************************************************************

The probability amplitudes $c_{k}(t)$ ($k=1,2,3$) of the three states satisfy the Schr\"{o}dinger equation that reads, in the rotating-wave approximation \cite{Allen-Eberly,Shore}, as
\begin{equation}
i\hbar \partial _{t}\mathbf{c}(t)=\mathbf{H}(t)\mathbf{c}(t),  \label{SEq}
\end{equation}%
where $\mathbf{c}(t)=[c_{1}(t),c_{2}(t),c_{3}(t)]^{T}$ \ and the Hamiltonian
is
\begin{equation}
\mathbf{H}(t)=\frac{\hbar }{2}\left[
\begin{array}{ccc}
0 & \Omega _{p}(t) & 0 \\
\Omega _{p}(t) & \Delta -i\Gamma & \Omega _{s}(t) \\
0 & \Omega _{s}(t) & 0%
\end{array}%
\right] .  \label{H}
\end{equation}%
Here $\Omega _{p}(t)=-\mathbf{d}_{12}\cdot \mathbf{E}_{p}(t)$ and $\Omega_{s}(t)=-\mathbf{d}_{32}\cdot \mathbf{E}_{s}(t)$ are the Rabi frequencies of the pump and Stokes pulses, respectively.
Each of them is proportional to the electric-field amplitude $\mathbf{E}_{p}(t)$ or $\mathbf{E}_{s}(t)$ of
the respective laser field and the corresponding transition dipole moment $\mathbf{d}_{12}$ or $\mathbf{d}_{32}$. 
The positive constant $\Gamma $ represents the irreversible loss rate from the intermediate state $|2\rangle $, and the frequency offset from resonance is measured by the single-photon detuning $\Delta $. 
To keep things simple both $\Omega _{p}(t)$ and $\Omega _{s}(t)$ will be considered to be real and positive, since their phases may be removed by redefining the probability amplitudes. 
Moreover, we assume that the Rabi frequencies have the same time dependence $f(t)$,
\begin{equation}
\Omega _{p,s}(t)=\Omega _{p,s}^{0}f(t).  \label{the same time dependence}
\end{equation}%
We assume that at the initial time $t_{i}$ the three-state system is in its
ground state $\left\vert 1\right\rangle $,
\begin{equation}\label{initial time}
c_{1}(t_{i}) =1, \quad c_{2}(t_{i}) =0, \quad
c_{3}(t_{i}) =0.
\end{equation}

\section{Effective two-state system: bright and dark states}

The dynamics of a three-state system on two-photon resonance is easily understood in the so-called bright-dark basis, which is composed of the states
\begin{subequations}
\begin{eqnarray}
\left\vert b\right\rangle &=&\sin \theta \left\vert 1\right\rangle +\cos
\theta \left\vert 3\right\rangle , \\
\left\vert d\right\rangle &=&\cos \theta \left\vert 1\right\rangle -\sin
\theta \left\vert 3\right\rangle , \\
\left\vert 2\right\rangle &=&\left\vert 2\right\rangle ,
\end{eqnarray}%
\end{subequations}
where
\begin{equation}
\tan \theta =\frac{\Omega _{p}(t)}{\Omega _{s}(t)}.
\end{equation}%
Unlike the situation of delayed pulses employed in STIRAP, here $\theta $ is constant because the pulses have the same time dependence, and so are the states $\left\vert b\right\rangle $ and $\left\vert d\right\rangle $. 
The transformation from the basis $\{\ket{1},\ket{2},\ket{3}\}$ to the basis $\{\ket{b},\ket{2},\ket{d}\}$ is given in the matrix form as
\begin{equation}
\left[\begin{array}{c}
c_{1}(t) \\
c_{2}(t) \\
c_{3}(t)%
\end{array}\right] \mathbf{=}\left[
\begin{array}{ccc}
\sin \theta & 0 & \cos \theta \\
0 & 1 & 0 \\
\cos \theta & 0 & -\sin \theta%
\end{array}%
\right] \left[
\begin{array}{c}
c_{b}(t) \\
c_{2}(t) \\
c_{d}(t)%
\end{array}%
\right] .  \label{transformation}
\end{equation}%
%here $\left\vert b\right\rangle $, $\left\vert 2\right\rangle $ and $\left\vert d\right\rangle $ states make up the basis, 
It casts equation (\ref{SEq}) into the equation
\begin{equation}
i\frac{d}{dt}\left[
\begin{array}{c}
c_{b}(t) \\
c_{2}(t) \\
c_{d}(t)%
\end{array}%
\right] \mathbf{=}\left[
\begin{array}{ccc}
0 & \Omega & -i\dot{\theta} \\
\Omega & \Delta -i\Gamma & 0 \\
i\dot{\theta} & 0 & 0%
\end{array}%
\right] \left[
\begin{array}{c}
c_{b}(t) \\
c_{2}(t) \\
c_{d}(t)%
\end{array}%
\right] ,  \label{DB}
\end{equation}%
where an overdot denotes a time derivative and the effective coupling $\Omega $ is given as
\begin{equation}
\Omega (t)=\sqrt{\Omega _{p}^{2}(t)+\Omega _{s}^{2}(t)}.
\label{effective coupling}
\end{equation}%
Equation (\ref{DB}) is true for generally time-dependent $\theta $, however, in the present case (\ref{the same time dependence}), $\theta $ is constant and $\dot{\theta}=0$. 
Consequently, similar to the situation of constant Rabi frequencies, the dark state $\left\vert d\right\rangle $ is decoupled
from states $\left\vert b\right\rangle $, $\left\vert 2\right\rangle $ and the three-state problem is reduced to a two-state problem involving states $\left\vert b\right\rangle $ and $\left\vert 2\right\rangle $ only, 
\begin{equation}
i\frac{d}{dt}\left[\begin{array}{c}
c_{b}(t) \\
c_{2}(t)%
\end{array}\right] = \left[
\begin{array}{cc}
0 & \Omega \\
\Omega & \Delta -i\Gamma%
\end{array}%
\right] \left[
\begin{array}{c}
c_{b}(t) \\
c_{2}(t)%
\end{array}%
\right] .  \label{effective two-stay system}
\end{equation}%
Here state $\left\vert 2\right\rangle $ is coupled to state $\left\vert b\right\rangle $ by $\Omega $.
It is off resonance by a detuning $\Delta $ and decays irreversibly out of the system at the rate $\Gamma $. 
It should be noted that the bright state $\left\vert b\right\rangle $ does not decay itself but is coupled to the decaying state $\left\vert 2\right\rangle $. 
Because the dark state $\left\vert d\right\rangle $ in equations (\ref{DB}) is decoupled from states $\left\vert b\right\rangle
$ and $\left\vert 2\right\rangle $, then its population $P_{d}(t)=\left\vert c_{d}(t)\right\vert ^{2}$ is preserved,
\begin{equation}
P_{d}(t)=P_{d}(t_{i})\mathbf{=}\left\vert c_{1}(t_{i})\cos \theta
-c_{3}(t_{i})\sin \theta \right\vert ^{2}=\cos ^{2}\theta ,
\label{dark state population}
\end{equation}%
where we have taken into account the assumed initial conditions (\ref{initial time}).

\section{Creation of superpositions}

In the strong decay regime ($\Gamma \gg \Omega _{p,s}$) all population in the effective two-state system (\ref{effective two-stay system}) will be lost, but the population in the dark state will be conserved (\ref{dark state population}). 
Therefore, this scenario will create a superposition between states $\left\vert 1\right\rangle $ and $\left\vert 3\right\rangle $, which can be easily derived from Eq. (\ref{transformation}),
\begin{subequations}
\label{supperpositions}
\begin{eqnarray}
c_{1}(t_{f}) &\mathbf{=}&c_{d}(t_{f})\cos \theta =\cos ^{2}\theta , \\
c_{3}(t_{f}) &\mathbf{=}&- c_{d}(t_{f})\sin \theta=-\sin \theta \cos \theta .
\end{eqnarray}
\end{subequations}

In principle, many superpositions are possible but we will examine numerically in more detail the most interesting case of the equal superposition between states $\left\vert 1\right\rangle $ and $\left\vert 3\right\rangle $, which is achieved when $\Omega _{p}=\Omega _{s}$, or equivalently, when $\theta =\pi /4$ (see Eq.(\ref{supperpositions})). 
In this case, states $\left\vert 1\right\rangle $ and $\left\vert 3\right\rangle $ will carry each $\frac14$ of the initial population and the remaining 50\% of the population will be lost. 
In the following, we analyze the required range of parameters by means of numerical calculation of Eq.~(\ref{SEq}).

%***************************************************************
\begin{figure}[tbh]
\centerline{\includegraphics[width=1\columnwidth]{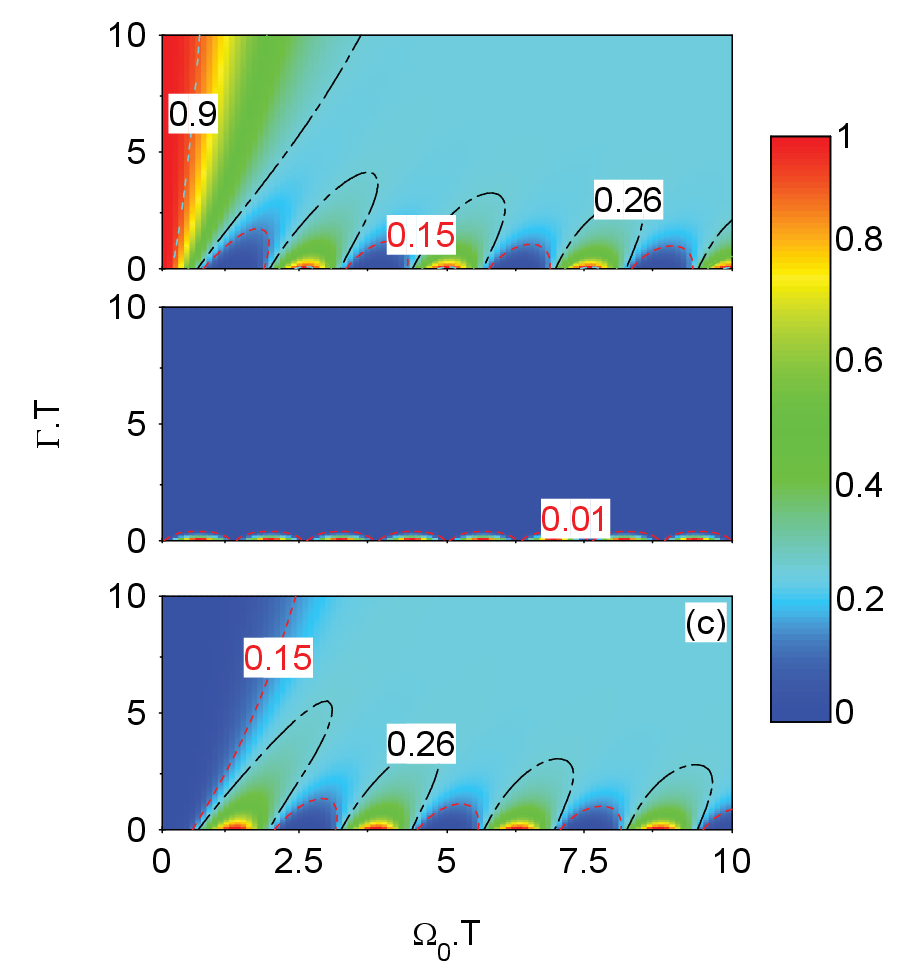}}
\caption{(Color online) Contour plot of the final populations at states $%
\left\vert 1\right\rangle $, $\left\vert 2\right\rangle $ and $\left\vert
3\right\rangle $ (Frames (a), (b) and (c) correspondingly) as obtained from
an integration of Eq. (\protect\ref{H}) as a function of the products $%
\Omega _{0}T$ and $\Gamma T$, with $T$ being the time duration of the
Gaussian pulses. In the big top-right areas in frames (a) and (c), one has $%
\left\vert c_{1}(t_{f})\right\vert ^{2}\approx \left\vert
c_{3}(t_{f})\right\vert ^{2}\approx 0.25$. Here the three-state quantum
system is on resonance ($\Delta =0$). In order to highlight better the
region where populations of states $\left\vert 1\right\rangle $ and $%
\left\vert 3\right\rangle $ are near 1/4, and the contour lines are chosen to be
not equidistant.}
\label{Contourplot}
\end{figure}
%***************************************************************

Figure \ref{Contourplot} shows the landscape of the populations of all three quantum states at the end of the evolution. The numerics are calculated with Eq. (\ref{H}) under the assumption of Gaussian-shaped pump and Stokes laser pulses,
\begin{subequations}
\label{Gaussian}
\begin{eqnarray}
\Omega _{p}(t) &=&\Omega _{0}\exp \left[ -\left(t/T\right) ^{2}%
\right] , \\
\Omega _{s}(t) &=&\Omega _{0}\exp \left[ -\left( t/T\right) ^{2}%
\right] ,
\end{eqnarray}%
\end{subequations}
and constant decay rate from the middle state $\Gamma $. 
For weak decay, the population oscillates (Rabi-like oscillations) between states $\left\vert 1\right\rangle $ and $\left\vert 3\right\rangle $, as seen on the bottom parts of Figs. \ref{Contourplot}(a) and \ref{Contourplot}(c). 
In this regime, the dynamics is not robust as the output depends strongly on the value of the coupling constant, as well as on the pulse duration $T$. 
In contrast, for large values of the products $\Omega _{0}T$ and $\Gamma T$ (upper right regions in Figs. \ref{Contourplot}(a) and \ref{Contourplot}(c)), there is a vast flat land where the output population in states $\left\vert 1\right\rangle $ and $\left\vert 3\right\rangle $ is close to $\frac 14$, as predicted by our analysis above. 
In this region, the output distribution is virtually independent of variation of the coupling constant and the decay rate. 
Figure \ref{Contourplot} allows us to identify the boundaries of this robust region, which are about $\Omega_{0}>5/T$ and $\Gamma >5/T$. 
This sets the limits of the coupling and decay rate for the method to work. 
However, it is worth mentioning that in the extreme case of very large decay rate and small coupling ($\Gamma \gg \Omega _{0}$) the system dynamics is frozen and all the population stays in the initial state~$\left\vert 1\right\rangle $ (see Fig. \ref{Contourplot}(a), upper left corner). 
This is the overdamped case, which has been studied in the literature \cite{Shore2006}.

%***************************************************************
\begin{figure}[tb]
\centerline{\includegraphics[width=1\columnwidth]{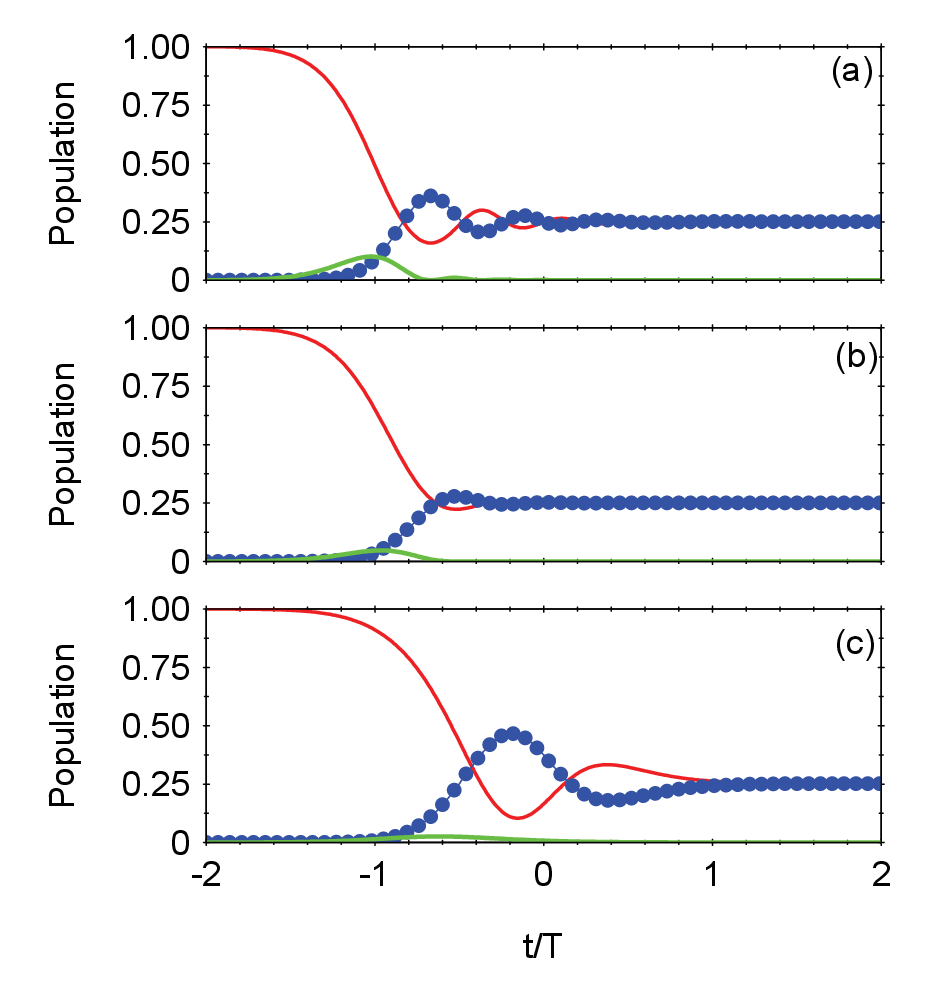}}
\caption{(Color online) Time evolution of all populations $P_{1}\left(
t\right) =\left\vert c_{1}\left( t\right) \right\vert ^{2}$ (solid red
lines), $P_{2}\left( t\right) =\left\vert c_{2}\left( t\right) \right\vert
^{2}$ (solid green lines), and $P_{3}\left( t\right) =\left\vert c_{3}\left(
t\right) \right\vert ^{2}$ (dotted blue lines) if initially all the
population is in state $\left\vert 1\right\rangle $, where the coupling in
all three frames is $\Omega _{0}=10/T$ , while the decay in frames (a) and
(c) is $\Gamma =10/T$ and in frame (b) is $\Gamma =5/T$. In frames (a) and
(b), the quantum system is in resonant (no detuning, $\Delta =0$), while in
(c) the detuning is $\Delta =25/T$. Again we assume the Gaussian shape Stokes
and pump laser pulses as Eq. (\protect\ref{Gaussian}).}
\label{evolution}
\end{figure}
%***************************************************************

The time evolution of all three populations $P_{1}\left( t\right) =\left\vert c_{1}\left( t\right) \right\vert ^{2}$ , $P_{2}\left( t\right) =\left\vert c_{2}\left( t\right) \right\vert ^{2}$, and $P_{3}\left( t\right) =\left\vert c_{3}\left( t\right) \right\vert ^{2}$ is depicted in Figs. \ref{evolution}(a) and \ref{evolution}(b) for the case when the quantum system is on resonance ($\Delta=0$). 
As seen in Fig. \ref{evolution}(a) for decay constant $\Gamma =5/T$, the behavior is characterized by damped oscillations between states $\left\vert 1\right\rangle $ and $\left\vert 3\right\rangle $ before reaching the stationary output superposition with $P_{1}=P_{3}=\frac14$. 
If the decay constant is doubled [Fig.~\ref{evolution}(b)], the oscillations are more strongly damped and the superposition state is reached earlier. 
Indeed, this superposition state is reached in both cases after a distance of about 10 times the amplitude decay time $1/\Gamma $. We have verified that for further increase of $\Gamma $ the above transient oscillations disappear completely and the dynamics become overdamped, as is the case for equivalent systems in classical or quantum physics \cite{Shore2006}.

We have already mentioned the remarkable feature that resonance is not required for the creation of an equal superposition between states $\left\vert 1\right\rangle $ and $\left\vert 3\right\rangle $. 
This is shown in Fig. \ref{evolution}(c), where the detuning is set to $\Delta =25/T$. 
It can be clearly recognized that the final state does not change despite the fact that the system is not resonant. 
Nevertheless, the transient oscillations become slower. 
Indeed, for a given value of the coupling constant $\Omega_{0}$ and the decay rate $\Gamma $, we have observed that an increase of $\Delta $ leads to stronger oscillatory character. 
Therefore, for very large $\Delta$ (as compared to $\Omega _{0}$ and $\Gamma $), a longer time is required before reaching the stationary final state.

Naively one can expect that because the time to create superpositions depends on the decay rate, then increasing the decay rate will give faster evolution.
However, one cannot expect faster evolution than the time needed for a single Rabi cycling between the bright state $\left\vert b\right\rangle $ and the excited state $\left\vert 2\right\rangle $.
Therefore the minimum time multiplied by the effective coupling (\ref{effective coupling}) should have the value of $\pi$, e.g., the minimum pulse area for exciting all the population from the bright state $\left\vert b\right\rangle $ to the excited state $\left\vert 2\right\rangle $.

\section{Discussions and Conclusions}

We presented a method for creation of coherent superpositions between the two ground states in a three-state quantum system in the $\Lambda$ configuration. 
Our approach uses the decay of the intermediate state as a crucial tool for the system to evolve and create superpositions. 
It uses the depletion of the bright state of the system by the population loss channel, which leaves the system in the dark state; the latter can be designed to be equal to any desired coherent superposition of the two ground states of the $\Lambda$ system.
We demonstrated that this mechanism is quite robust and effective by numerical simulations. 
As long as the losses from the intermediate state are sufficiently large, the method can operate on a short time scale. 
The method can be seen as an alternative to the famous technique STIRAP, which operates in a similar system.
STIRAP aims at eliminating the effect of the decay of the intermediate state by enforcing adiabatic evolution; the latter requires large pulse areas, i.e. large interaction times. 
When the intermediate state is very lossy for a given time duration, one can opt for the present technique, which benetifs from the strong decay, rather than prolonging the process until the adiabatic conditions are met. 
The price to pay is that $\frac12$ of the initial population is lost and the remaining $\frac12$ of the population is distributed to create an equal superposition ($\frac14:\frac14$) between the two ground states of the system.
This problem can be removed in an experiment using post-selection, e.g., by monitoring the population loss channel (which generate photoionization electrons, light-induced fluorescence, etc.).
Because in one-half of the events either a coherent superposition will be created or the population will be lost (but not both simultaneously), only the events when the loss channel fails to produce a signal will be accounted for.

Besides the application for creating equal superpositions in quantum physics, the current technique could be applied in other areas of physics, in particular in waveguide optics \cite{Alrifai} and nonlinear optics \cite{Al-Mahmoud} where direct analogs with Schr\"{o}dinger equation exist \cite{Longhi,Suchowski}.

%----------------------------------------
\acknowledgements%----------------------------------------

This research is partially supported by the Bulgarian national plan for recovery
and resilience, contract BG-RRP-2.004-0008-C01 SUMMIT: Sofia University
Marking Momentum for Innovation and Technological Transfer, project number
3.1.4.

\end{document}